\documentclass[12pt]{article}
\usepackage{amsfonts}
\usepackage{amssymb}
\usepackage{amsbsy}
\usepackage{graphics}
\usepackage{epsfig}

\input epsf.sty

\newlength{\TZ}
\newcommand{\BEQ}{\begin{equation}}     
\newcommand{\BEA}{\begin{eqnarray}}
\newcommand{\BD}{\begin{displaymath}}
\newcommand{\EEQ}{\end{equation}}       
\newcommand{\EEA}{\end{eqnarray}}
\newcommand{\ED}{\end{displaymath}}
\newcommand{\eps}{\varepsilon}          
\newcommand{\D}{{\rm d}}                
\newcommand{\II}{{\rm i}}               
\newcommand{\sign}{{\rm sign\,}}        
\newcommand{\demi}{\frac{1}{2}}         
\newcommand{\wit}[1]{\widetilde{#1}}    
\newcommand{\wht}[1]{\widehat{#1}}      

\renewcommand{\vec}[1]{\boldsymbol{#1}} 

\newcommand{\vekz}[2]
     {\mbox{${\begin{array}{c} #1  \\ #2 \end{array}}$}}
\newcommand{\matz}[4] 
     {\mbox{${\begin{array}{cc} #1 & #2 \\ #3 & #4 \end{array}}$}}
     
\catcode`\@=11
\def\numberbysection{\@addtoreset{equation}{section}
        \def\theequation{\thesection.\arabic{equation}}}
\numberbysection

\begin{document}

\begin{titlepage}

\vskip 1.5 cm
\begin{center}
{\LARGE \bf Causality from dynamical symmetry: 
an example from local scale-invariance\protect{\footnote{Invited talk presented 
at conference `Algebra Geometry Mathematical Physics' AGMP-7 in Mulhouse (France), 
24$^{\rm th}$-26$^{\rm th}$ of October 2011}}
}
\end{center}

\vskip 2.0 cm
\centerline{{\bf Malte Henkel}}
\vskip 0.5 cm
\centerline{Groupe de Physique Statistique,
D\'epartement de Physique de la Mati\`ere et des Mat\'eriaux,}
\centerline{Institut Jean Lamour,\footnote{Laboratoire associ\'e 
au CNRS UMR 7198} 
Universit\'e de Lorraine Nancy,} 
\centerline{B.P. 70239, F -- 54506 Vand{\oe}uvre l\`es Nancy Cedex, France}

\begin{abstract}
Physical ageing phenomena far from equilibrium naturally lead to dynamical scaling. 
It has been proposed to consider the
consequences of an extension to a larger Lie algebra of local scale-transformation. 
The best-tested applications of this
are explicitly computed co-variant two-point functions 
which have been compared to non-equilibrium response functions in a
large variety of statistical mechanics models. 
It is shown that the extension of the Schr\"odinger Lie algebra
$\mathfrak{sch}(1)$ to a maximal parabolic sub-algebra, 
when combined with a dualisation approach, 
is sufficient to derive the causality condition required for the
interpretation of two-point functions as physical response functions. 
The proof is presented for the recent logarithmic
extension of the differential operator representation of the Schr\"odinger algebra. 
\end{abstract}

\end{titlepage}

\setcounter{footnote}{0} 

\section{Motivation and background}

Physicists have valued since a long time the important r\^ole of symmetries, be it
for their usefulness in simplifying practical calculations, be it for making progress
in issues of conceptual understanding. Arguably the most famous instance of this is
{\em relativistic covariance} in mechanics and electrodynamics,\footnote{Physicists carefully distinguish between
{\em co-}variance and {\em in}variance: for example, a scalar is invariant under rotations, while a vector or a tensor
transforms covariantly. Since the equations of mechanics or electrodynamics are in general vector or tensor equations, 
it is appropriate to speak of relativistic co-variance.}  formally described by the
Lie group of Lorentz transformations which has been introduced almost exactly a 
century ago \cite{Lorentz04,Einstein05}. Almost three quarters of a century later, 
it has been realised that by the inclusion of scale-invariance and the subsequent extension
of the Lorentz group to the {\em conformal group} 
considerable advances can be made, simultaneously in
cooperative phenomena in statistical mechanics as well as in string theory.
A special r\^ole is herein played by the case of two dimensions, where the infinite-dimensional
Lie algebra of conformal transformations is centrally extended to the Virasoro algebra, in order
to be able to take the physical effects 
of either thermal or quantum fluctuations into account \cite{Belavin84}. 

Here, we shall consider a different example of covariance under 
a certain class of space-time transformations. 
Historically, these were found by considering 
the dynamical symmetries of what in physics is called
by an abuse of language the `non-relativistic limit' 
of mechanics where the speed of light $c\to\infty$. 
Specifically, we shall be interested in the transformations of the 
{\em Schr\"odinger group} {\sl Sch}($d$)
which is defined by the following transformation on space-time coordinates 
$(t,\vec{r})\in\mathbb{R}\times\mathbb{R}^d$:
\BEQ
t \mapsto t' := \frac{\alpha t +\beta}{\gamma t + \delta} \;\; , \;\;
\vec{r} \mapsto \vec{r}' := \frac{{\cal R}\vec{r} + \vec{v}t +\vec{a}}{\gamma t +\delta} 
\;\; ; \;\; \alpha\delta-\beta\gamma=1
\EEQ
with ${\cal R}\in{\sl SO}(d)$, $\vec{a},\vec{v}\in\mathbb{R}^d$ and 
$\alpha,\beta,\gamma,\delta\in\mathbb{R}$. 
Indeed, it has been known to mathematicians since a long 
time that free-particle motion (be it classical, quantum
mechanical or probabilistic) is invariant under the 
Schr\"odinger group in the sense that a solution of the
equation of motion is mapped onto a different solution of 
the same equation of motion in the transformed coordinates
\cite{Jacobi1843,Lie1881}. During the past century, 
this has been re-discovered a couple of times, both in mathematics
and physics, see e.g. \cite{Henkel10} and references therein. 
It is often convenient to study instead the Lie algebra 
$\mathfrak{sch}(d) = \mbox{\rm Lie}(\mbox{\sl Sch}(d)) 
= \left\langle X_{0,\pm 1}, Y_{\pm 1/2}^{(j)}, M_0, R_0^{(jk)}\right\rangle_{j,k=1,\ldots d}$ 
with the explicit generators (where $\partial_j := \partial/\partial r_j$ and 
$\vec{\nabla}_{\vec{r}} = (\partial_1, \ldots, \partial_d)^{\rm T}$)
\BEA
X_n &=& -t^{n+1}\partial_t - \frac{n+1}{2}t^n \vec{r}\cdot\vec{\nabla}_{\vec{r}} 
- \frac{\cal M}{2}(n+1)n t^{n-1} \vec{r}^2
- \frac{n+1}{2} x t^n \nonumber \\
Y_m^{(j)} &=& - t^{m+1/2} \partial_j - \left( m + \demi\right)  t^{m-1/2} {\cal M} r_j \nonumber \\
M_n &=& - t^n {\cal M} \label{1.5} \\
R_n^{(jk)} &=& - t^n \bigl( r_j \partial_k - r_k \partial_j \bigr) \:=\: - R_n^{(kj)} \nonumber
\EEA
Herein, the non-derivative terms (characterised by a dimensionful constant $\cal M$ (`mass') 
and a scaling dimension $x$) describe
how the solution of a Schr\"odinger/diffusion equation will transform under the action of 
$\mathfrak{sch}(d)$. One has the
non-vanishing commutation relations  
\newpage\typeout{*** saut de page ***}
\BEA
{} \bigl[ X_n, X_{n'}\bigr] &=& (n-n') X_{n+n'} \hspace{1.2truecm} \;\; , \;\; \hspace{0.33truecm}
{} \bigl[ X_n, Y_m^{(j)}\bigr] \:=\: \left(\frac{n}{2} -m\right) Y_{n+m}^{(j)}
\nonumber \\
{} \bigl[ X_n, M_{n'} \bigr] &=& - n' M_{n+n'} \hspace{1.9truecm}\;\; , \;\; \hspace{0.33truecm}
{} \bigl[ X_n, R_{n'}^{(jk)} \bigr] \:=\: -n' R_{n+n'}^{(jk)} \nonumber \\
{} \bigl[ Y_{m}^{(j)}, Y_{m'}^{(k)} \bigr] &=& \delta^{j,k}\,
\left (m - m'\right) M_{m+m'}  \;\; , \;\; 
\hspace{0.3truecm} \bigl[ R_n^{(jk)},Y_m^{(\ell)} \bigr] 
\:=\: \delta^{j,\ell}\, Y_{n+m}^{(k)} -
\delta^{k,\ell}\, Y_{n+m}^{(j)} \label{gl:schcr} 
\EEA 
up to the commutators of $\mathfrak{so}(d)$, which are not spelled out. 
The Schr\"odinger algebra is also the Lie symmetry algebra of non-linear (systems of)
equations. Probably one of the best-known examples of this kind are the Euler equations of a compressible fluid
of velocity $\vec{u}=\vec{u}(t,\vec{r})$ and density $\rho=\rho(t,\vec{r})$
\BEQ
\partial_t \rho + \vec{\nabla}\cdot\bigl( \rho \vec{u}\bigr) =0 \;\; , \;\;
\rho \bigl( \partial_t + (\vec{u}\cdot\vec{\nabla})\bigr)\vec{u} +
\vec{\nabla} P =0 
\label{gl:1.4}
\EEQ 
together with the polytropic equation of state $P = \rho^{1+2/d}$. This has been known
to russian and ukrainian mathematicians at least since the 1960s \cite{Ovsiannikov80,Fush93} 
and was re-discovered by
european physicists around the turn of the century \cite{Hassaine00,ORaif01}. 
Many more Schr\"odinger-invariant non-linear equations and systems exist, see \cite{Fush89,Fush93,Fush95,Rideau93}. 
Analogously to conformal invariance
in $2D$, an infinite-dimensional extension of $\mathfrak{sch}(d)$ is the 
{\em Schr\"odinger-Virasoro algebra} 
$\mathfrak{sv}(d) = \left\langle X_n, Y_m^{(j)}, M_n, 
R_n^{(jk)}\right\rangle_{n\in\mathbb{Z},m\in\mathbb{Z}+\demi,j,k\in\{1,\ldots,d\}}$, 
with an explicit representation
in (\ref{1.5}) and an immediate extension of the commutators (\ref{gl:schcr}) \cite{Henkel94}. 
The mathematical properties of $\mathfrak{sv}$ are studied in detail in \cite{Roger06,Unterberger11},
the geometry in \cite{Duval09}  
and physical applications are reviewed in \cite{Henkel10}. 

Contrary to a widespread belief, when taking the non-relativistic 
limit of the conformal algebra, one does {\em not}
obtain the Schr\"odinger algebra, but a different Lie algebra, which by now is usually  called
the {\em conformal Galilean algebra} 
$\mbox{\sc cga}(d)=\langle X_{\pm 1,0},Y_{\pm 1,0}^{(j)},R_0^{(jk)}\rangle_{j,k=1,\ldots,d}$ 
\cite{Havas78,Henkel97,Negro97,Henkel03a,Bagchi09,Martelli09}. 
Its most general known differential operator representation is \cite{Cherniha10}
\BEA X_n &=&
- t^{n+1}\partial_t - (n+1) t^n \vec{r}\cdot\vec{\nabla}_{\vec{r}}
 - n(n+1) t^{n-1} \vec{\gamma}\cdot\vec{r} - x (n+1)t^n
\nonumber \\
Y_n^{(j)} &=& - t^{n+1} \partial_{j} - (n+1) t^n \gamma_j  \label{1.6} \\
R_n^{(jk)} &=& - t^n \bigl( r_j \partial_{k} -  r_k \partial_{j} \bigr)
- t^n \bigl( \gamma_j \partial_{\gamma_k}-\gamma_k
\partial_{\gamma_j}\bigr) \:=\: - R_n^{(kj)} \nonumber 
\EEA
where $\vec{\gamma}=(\gamma_1,\ldots,\gamma_d)$ is a vector of dimensionful constants and 
$x$ is again a scaling dimension. 
Its non-vanishing commutators read, again up to those of $\mathfrak{so}(d)$
\BEA 
{}[X_n, X_{n'}] &=& (n-n') X_{n+n'} \;\;,\;\; 
{}[X_n, Y_{m}^{(j)}] \:=\: \left(n-m\right)Y_{n+m}^{(j)} \nonumber \\
{}[X_n, R_{n'}^{(jk)}] &=& -n' R_{n+n'}^{(jk)} \hspace{0.75truecm}\;\; , \;\; 
{}[R_n^{(jk)}, Y_m^{(\ell)}] 
\:=\: \delta^{j,\ell} Y_{n+m}^{(k)} - \delta^{k,\ell} Y_{n+m}^{(j)} \label{1.7}
\EEA
The non-linear systems for which $\mbox{\sc cga}(d)$ arises as a 
(conditional) dynamical symmetry are distinct from (\ref{gl:1.4}) \cite{Zhang10,Cherniha10}.
As before, the systematic organisation of the generators allows 
for an immediate infinite-dimensional extension 
$\mathfrak{av}(d) := \left\langle X_n, Y_n^{(j)}, R_n^{(jk)}\right\rangle_{n\in\mathbb{Z},j,k=1,\ldots d}$ 
\cite{Henkel97,Ovsienko98} ({\it `altern-Virasoro algebra'}).

In $d=2$ spatial dimensions, it was recently shown \cite{Lukierski06} 
that the conformal Galilean algebra admits a
so-called `exotic' central extension. This is achieved by adding to
the commutator relations (\ref{1.7}) the following commutator 
\BEQ
\label{1-6} {} \bigl[ Y_n^{(1)}, Y_m^{(2)} \bigr] \:=\:
\delta_{n+m,0}\, \bigl( 3\delta_{n,0} -2 \bigr)\, \Theta, \quad
n,m\in\{\pm 1,0\}, 
\EEQ 
where the new central generator $\Theta$ is
needed for this central extension. Physicists usually call this
central extension of {\sc cga}$(2)$ the {\it exotic Galilean
conformal algebra}, and we shall denote it by 
$\mbox{\sc ecga} = \mbox{\sc cga}(2) + \mathbb{C} \Theta$. 
A differential operator representation of {\sc ecga} reads \cite{Martelli09,Cherniha10} 
\BEA
{} X_n &=& -t^{n+1}\partial_t - (n+1) t^n \vec{r}\cdot\vec{\nabla}_{\vec{r}} -
\lambda(n+1) t^n  - (n+1)n t^{n-1} \vec{\gamma}\cdot\vec{r} - (n+1)n
\vec{h}\cdot\vec{r}
\nonumber \\
Y_n^{(j)} &=& - t^{n+1}\partial_{j} - (n+1) t^n \gamma_j - (n+1) t^n
h_j - (n+1) n  \left(r_2 - r_1 \right) \theta \label{gl:exaltrep} \\
R_0^{(12)} &=& - \bigl( r_1 \partial_{2} -  r_2 \partial_{1} \bigr)
- \bigl( \gamma_1 \partial_{\gamma_2}  -
\gamma_2\partial_{\gamma_1}\bigr) - \frac{1}{2\theta} \vec{h}\cdot
\vec{h} \nonumber
\EEA
where $n\in\{\pm 1,0\}$  and 
$j,k\in\{1,2\}$.\footnote{An infinite-dimensional extension of 
{\sc ecga} does not appear to be possible.} 
Because of Schur's lemma, the central generator $\Theta$ can be
replaced by its eigenvalue $\theta\ne 0$. The components of the
vector-operator $\vec{h}=(h_1,h_2)$ are connected by the
commutator $[h_1, h_2] = \Theta$. For illustration, we quote the following non-linear system
which has {\sc ecga} as a Lie symmetry \cite{Cherniha10}
\BEQ
\vec{\nabla} \wedge \vec{u} = \vec{0} \;\; , \;\; 
\partial_t \vec{u} + \left(\vec{u}\cdot\vec{\nabla}\right)\vec{u} 
+\demi \left( \vec{u}\wedge \vec{\nabla}\right)\wedge \vec{u} 
= q \vec{\nabla} \wedge \vec{\omega}
\label{gl:1.9}
\EEQ
where $q$ is a constant, $\vec{u}=\vec{u}(t,\vec{r})= (u_1, u_2,0)^{\rm T}$  
is a planar vector embedded into $\mathbb{R}^3$ (and similarly for $\vec{\nabla}$) and 
$\vec{\omega}=(0,0,w)^{\rm T}$ is constructed from the coordinate dual to the
central charge according to $\Theta=\partial_w$. Clearly, (\ref{gl:1.9}) is very different
from (\ref{gl:1.4}). \\

\noindent
{\bf Remark:} In analogy to the Virasoro algebra of $2D$ conformal invariance, it is natural to
ask if the full definition of algebras such as $\mathfrak{sv}(d)$ or $\mathfrak{av}(d)$ may include
central extensions. For the Schr\"odinger-Virasoro algebra $\mathfrak{sv}(1)$, one merely has the
central Virasoro-like extension of $[X_n,X_m]$ \cite{Henkel94,Roger06,Unterberger11}. On the other hand,
if in $\mathfrak{sv}(1)$ one considers the generators $Y_n$ with {\em integer} indices $n\in\mathbb{Z}$, 
then three distinct central extensions are possible \cite{Roger06}, \cite[Thm 7.4]{Unterberger11}. 
Finally, for the `altern-Virasoro algebra' or the infinite-dimensional extension of $\mbox{\sc cga}(1)$
one has the central extensions \cite{Ovsienko98,Henkel06b}  
\BEQ \label{1.10}
{} \left[ X_n, X_{n'} \right] = (n-n') X_{n+n'} + \frac{c_X}{12} \delta_{n+n',0} \left( n^3 - n\right) 
\;\; , \;\;
{} \left[ X_n, Y_{n'} \right] = (n-n') X_{n+n'} + \frac{c_Y}{12} \delta_{n+n',0} \left( n^3 - n\right) 
\EEQ
with two independent central charges. The independence of the two central charges $c_{X,Y}$ can be
illustrated by the following example: let $L_n$ and $L_n'$ with $n\in\mathbb{Z}$ stand for the
generators of two commuting Virasoro algebras with central charges $c$ and $c'$. Then the generators 
\BEQ
X_n := \left(\matz{L_n+L_n'}{0}{0}{L_n+L_n'}\right) \;,\;
Y_n := \left(\matz{0}{L_n}{0}{0}\right) \;,\;
K_X := \left(\matz{1}{0}{0}{1}\right) \;,\;
K_Y := \left(\matz{0}{1}{0}{0}\right)
\EEQ
satisfy the commutators (\ref{1.10}), with $c_X=(c+c') K_X$ and $c_Y=c K_Y$ 
\cite{Henkel06b} \cite[Exerc.5.5]{Henkel10}. \\

In statistical physics, many situations are known and well-understood where the usual space-time
symmetries of temporal and spatial translation-invariance and rotation-invariance are supplemented
by dilatation (or scale-) invariance.\footnote{In the physicists terminology: at an equilibrium critical point, the
partition function is {\em in}variant under dilatations, 
whereas correlators of physical observables transform {\em co-}variantly.} 
The paradigmatic examples are provided by various
phase transitions -- often-mentioned examples include the liquid-gas transition, 
the ferromagnetic-paramagnetic
transition, the transition between normal conductivity and superconductivity, 
the electroweak phase transition
in the early universe and so on. Here, 
we shall be interested in instances of {\em dynamical} scaling, which
involves the space-time rescaling $t\mapsto b^{z} t$, 
$\vec{r}\mapsto b \vec{r}$ and is characterised by a
constant, the {\em dynamical exponent} $z$. It 
arises naturally in various many-body systems far from equilibrium, often without having to 
fine-tune external parameters. 
Paradigmatic examples are {\em ageing phenomena}, which may arise in systems quenched, from some
initial state, either (i) into a coexistence phase 
with more than one stable equilibrium state or else 
(ii) onto a critical point of the stationary state, 
see \cite{Bray94a,Cugliandolo02,Henkel10} for reviews. 
We shall adopt a phenomenological
point of view and characterise ageing through three (symmetry) properties: namely \cite{Henkel10}
\begin{enumerate}
\item slow, non-exponential relaxation, 
\item breaking of time-translation-invariance 
\item dynamical scaling.
\end{enumerate} 
For equilibrium critical phenomena, it was believed for a long time that under relatively weak 
conditions scale-invariance
could be extended to conformal invariance. Recent work 
has considerably clarified that this conclusion cannot
always be drawn so readily \cite{Riva05}, although there exist many theoretical models which are 
indeed both scale- and conformally invariant, with
many important consequences \cite{Polyakov70,Belavin84}. 
Drawing on this analogy, we look for situations 
when dynamical scaling can be extended to a larger group, such as the Schr\"odinger
group when $z=2$. Quite analogously with respect to conformal invariance, 
one is looking for co-variant two-point functions,
such that the co-variance under Schr\"odinger transformations 
leads to a set of differential equations for the said two-point
function.  However, in contrast to conformal invariance, 
it has turned out that this kind of co-variance condition is {\em not}
satisfied by correlation functions but rather by the so-called 
{\em response functions}. As an example, we quote the
basic prediction of Schr\"odinger-invariance for the linear
two-time auto-response function \cite{Henkel01b,Henkel02,Henkel03a,Henkel06a}
\BEA 
R(t,s) &=& \left.\frac{\delta \langle\phi(t,\vec{r})\rangle}{\delta h(s,\vec{r})}\right|_{h=0} 
\:=\: \left\langle \phi(t,\vec{r}) \wit{\phi}(s,\vec{r}) \right\rangle 
\:=\: s^{-1-a} f_R\left(\frac{t}{s}\right) \;\; , \;\; 
\nonumber \\
f_R(y) &=& f_0 y^{1+a'-\lambda_R/z} (y-1)^{-1-a'} \Theta(y-1)
\label{R}
\EEA
which measures the linear response of the order-parameter $\phi(t,\vec{r})$ 
with respect to its canonically conjugated external field $h(s,\vec{r})$. 
In stochastic field-theory using the
Janssen-de Dominicis  formalism, see e.g. \cite{Cugliandolo02,Henkel10}, 
it can be shown that response functions can be written as a correlator
between the order-parameter $\phi$ and an associated {\em `response field'} 
$\wit{\phi}$.\footnote{The example of the free field
equations of motion already shows that while the order-parameter $\phi$ 
has a positive `mass' ${\cal M}>0$, the `mass' associated to
the response field is negative\label{massereponse} $\wit{\cal M} = - {\cal M} <0$.} 
The auto-response exponent $\lambda_R$ and the ageing exponents $a,a'$ are universal
non-equilibrium exponents.\footnote{In magnets, with the temperature 
rapidly lowered (`quenched') from a very high initial value to
a finite value $T$, mean-field theory suggests that generically
$a=a'$ for quenches to low temperatures $T<T_c$ and $a\ne a'$ 
for critical quenches at $T=T_c$, where $T_c$ is the equilibrium critical
temperature \cite{Henkel10}.} 
This prediction has been tested extensively, 
and the computation of correlators can be understood along different lines, 
as reviewed in \cite{Henkel10}. 

\begin{figure}[tb]
\centerline{\psfig{figure=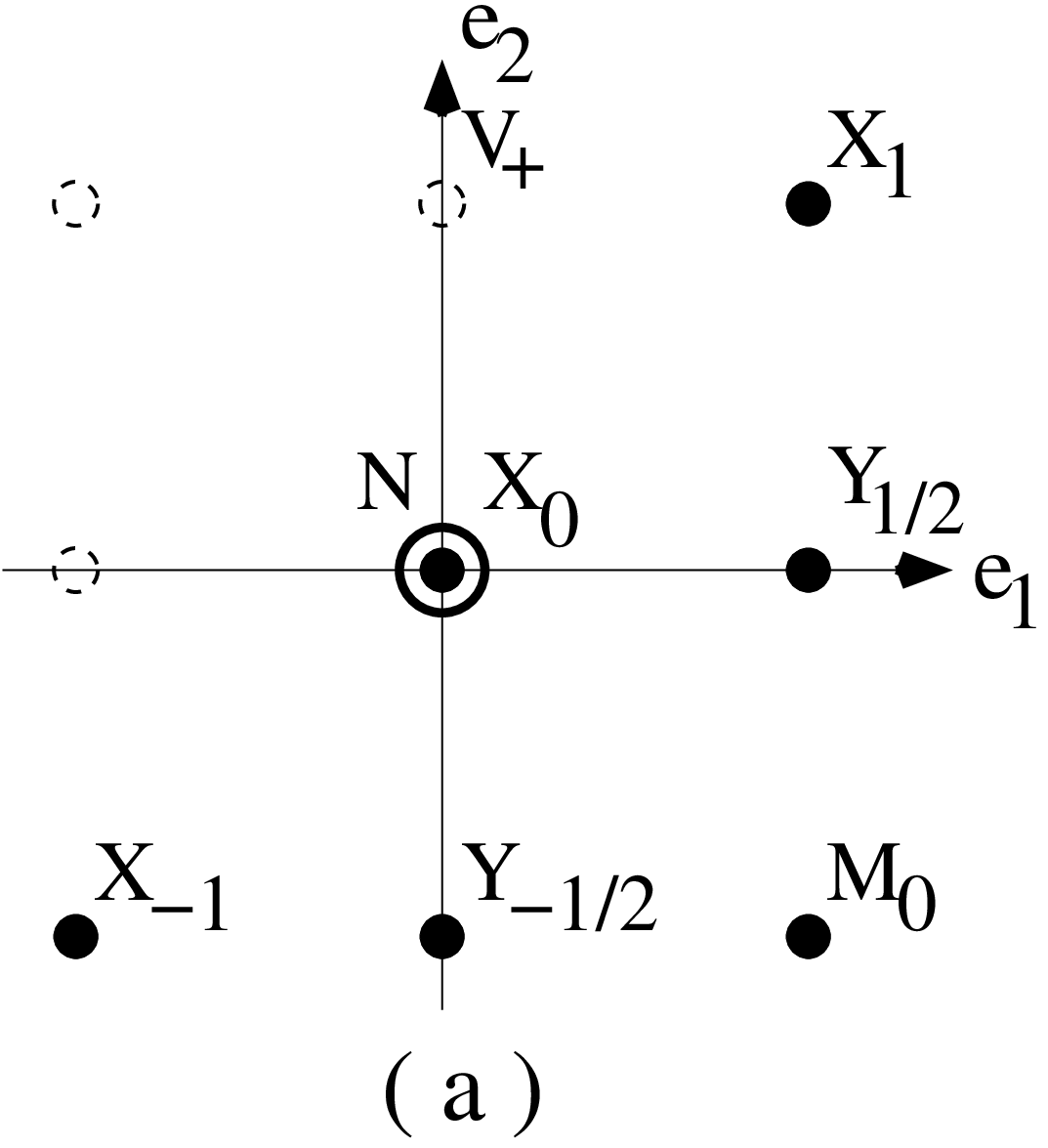,width=3.4in,clip=} ~~~~
\psfig{figure=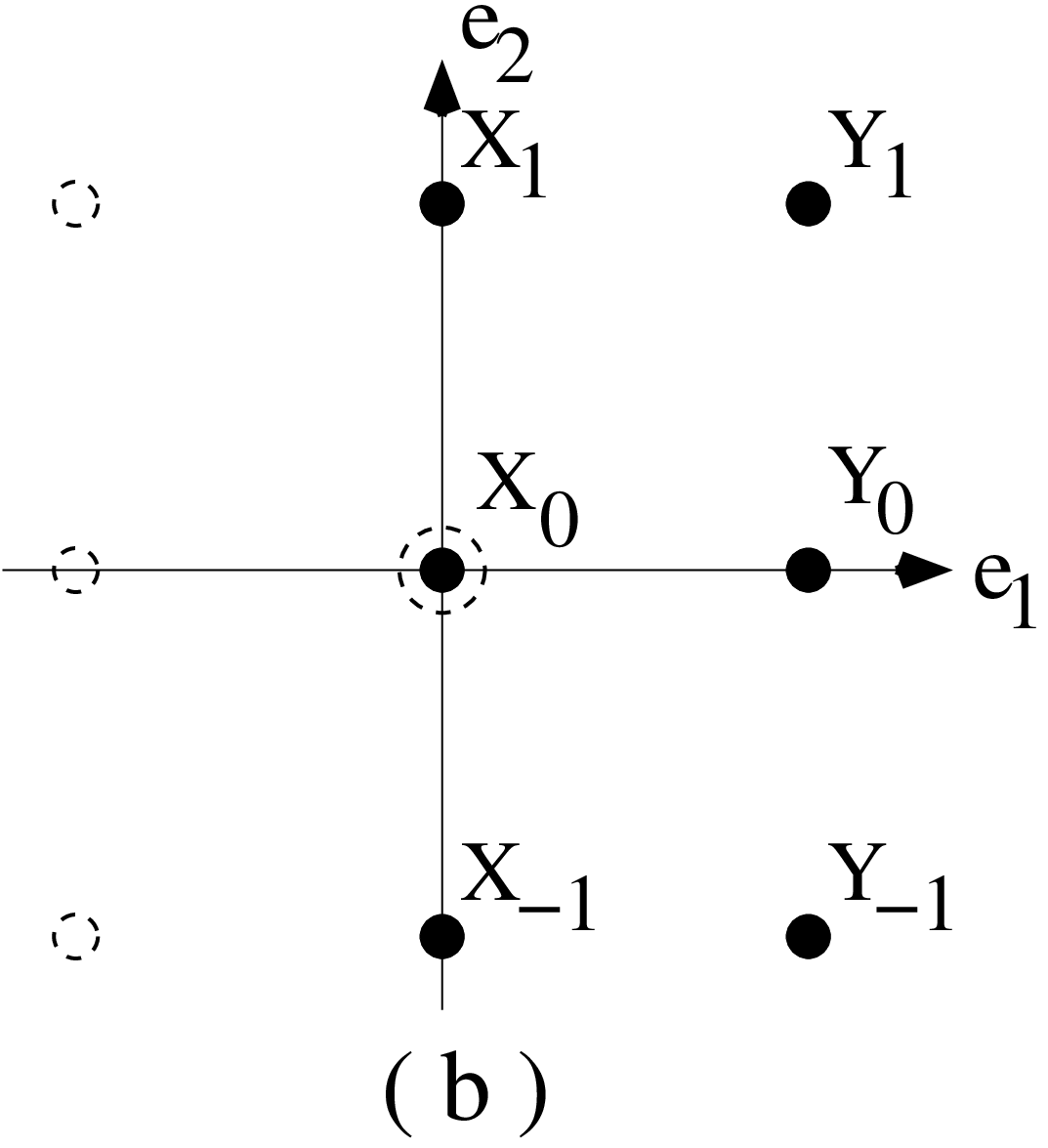,width=3.4in,clip=}}
\caption[figab]{\label{figab} Root diagrammes of some 
sub-algebras of the complex Lie algebra $B_2$. The roots
of $B_2$ are indicated by the full and broken dots, 
those of the sub-algebras by the full dots only. \\
(a) Schr\"odinger algebra 
$\mathfrak{sch}(1)=\left\langle X_{\pm 1,0}, Y_{\pm 1/2}, M_0\right\rangle$ and the
maximal parabolic sub-algebra 
$\wit{\mathfrak{sch}}(1)=\mathfrak{sch}(1)+\mathbb{C} N$. \\
(b) Conformal Galilean algebra 
$\mbox{\sc cga}(1) = \left\langle X_{\pm 1,0}, Y_{\pm 1,0}\right\rangle$. 
}
\end{figure}

The main distinction of response functions with respect 
to correlation functions is the {\em causality condition} $t>s$, which
is spelt out in (\ref{R}) through the Heaviside $\Theta$-function. 
Here, we shall show {\it how the origin of this causality condition
can be understood from an algebraic symmetry hypothesis}. 
The central observation is that there exists a natural way to imbed the
Schr\"odinger algebra $\mathfrak{sch}(d)$ 
into a (semi-simple) conformal Lie algebra in $d+2$ dimensions \cite{Burdet73a,Henkel03a}. 
This opens the route to introduce a powerful mathematical concept, 
namely the parabolic sub-algebras of that conformal
Lie algebra. By definition, a (standard) {\it parabolic sub-algebra} 
is made up by the Cartan sub-algebra and a selected set of
positive roots \cite{Knapp86}. It turns out that 
{\it a sufficient condition for deriving a causality condition for the
co-variant two-point functions as in (\ref{R}) 
is the co-variance under a maximal parabolic sub-algebra dualised in such a way that
translation-invariance in the dual variable becomes part of the algebra}. For example, 
rather than requiring Schr\"odinger-covariance under the algebra 
$\mathfrak{sch}(d)$, one considers an extended
co-variance under the {\em maximal parabolic sub-algebra} 
$\wit{\mathfrak{sch}}(d) = \mathfrak{sch}(d) + \mathbb{C} N$,
with a single extra generator $N$, to be specified below \cite{Henkel03a}. 
In figure~\ref{figab}a, we illustrate the inclusion,
for the $d=1$ case, $\wit{\mathfrak{sch}}(1)=\mathfrak{sch}(1)+\mathbb{C} N \subset B_2$ to the
complex conformal Lie algebra $B_2$, isomorphic to the conformal algebra $B_2 \stackrel{~}{=}\left(\mathfrak{conf}(3)\right)_{\mathbb{C}}$ in three dimensions. 
Similarly, figure~\ref{figab}b illustrates the inclusion 
$\mbox{\sc cga}(1)\subset B_2$ and an extension by the second independent generator in the Cartan sub-algebra
would give an inclusion $\wit{\mbox{\sc cga}}(1)\subset B_2$. Maximal parabolic sub-algebras of $B_2$ are distinguished
in that the addition of any further generator produces the entire conformal algebra. Furthermore, 
in view of may important physical applications (some of them to be mentioned briefly below),
we shall see that the same kind of causality 
condition is also obtained for the novel logarithmic extensions of the
Schr\"odinger and/or conformal Galilean algebras \cite{Hosseiny10,Henkel10b,Hosseiny11,Setare12,Hyun12,Hyun13}. 

This paper is organised as follows. 
The first sections recall basic facts on the ingredients required. 
In section~2, we recall briefly those elements of 
{\em logarithmic} conformal invariance as
required here and quote the corresponding logarithmic extensions of 
$\mathfrak{sch}(d)$- and $\mbox{\sc cga}(d)$-invariance. 
In section~3, specialising to $d=1$ for brevity, 
we describe the inclusion of the Schr\"odinger algebra into 
$B_2$ by a canonical dualisation procedure and its extension to the logarithmic
case. In section~4, the shapes of the dual logarithmic Schr\"odinger-covariant 
two-point functions will be derived and we shall
see that Schr\"odinger-covariance alone is {\em not} enough to derive a causality condition. 
In section~5 we finally derive our main result, namely that 
$\wit{\mathfrak{sch}}(1)$-covariant two-point functions 
automatically must obey causality. In this way, a combination of dualisation with an extended 
dynamical co-variance requirement allows to derive the causality condition algebraically.  

\section{Logarithmic conformal invariance} 

In various physical situations presenting an equilibrium phase transition, 
for example disordered systems \cite{Caux96}, 
percolation \cite{Flohr05,Mathieu07} or sand-pile models \cite{Poghosyan07}, 
it has been useful to consider degenerate vacuum states. Formally, 
this can be implemented \cite{Gurarie93,Rahimi97} by replacing the order parameter
$\phi$ by a vector $\left(\vekz{\psi}{\phi}\right)$ and the scaling
dimension $x$ by a Jordan matrix $\left(\matz{x}{1}{0}{x}\right)$. 
For reviews, see \cite{Flohr03,Gaberdiel03}.

Here, we consider an analogous extension of the 
representations of the Schr\"odinger and conformal Galilean algebras. 
Consider the two-point functions\footnote{Here and later, $\left\langle \cdot \right\rangle$ refers to an average over the
thermal noise.} 
\BEQ \label{2.1}
F := \left\langle \phi_1(t_1,\vec{r}_1) \phi_2(t_2,\vec{r}_2)\right\rangle \;\; , \;\; 
G := \left\langle \phi_1(t_1,\vec{r}_1) \psi_2(t_2,\vec{r}_2)\right\rangle \;\; , \;\; 
H := \left\langle \psi_1(t_1,\vec{r}_1) \psi_2(t_2,\vec{r}_2)\right\rangle
\EEQ 
Temporal and spatial translation-invariance imply that 
$F=F(t,\vec{r}), G=G(t,\vec{r})$ and $H=H(t,\vec{r})$ with $t=t_1-t_2$ and
$\vec{r}=\vec{r}_1-\vec{r}_2$. 
Since we shall explain the method in more detail below, we now simply quote the results and
generalise them immediately to an arbitrary space dimension $d$. 
Co-variance under the logarithmic extension of 
either $\mathfrak{sch}(d)$ or $\mbox{\sc cga}(d)$ implies $x_1=x_2 =: x$ and $F=0$. 
For logarithmic Schr\"odinger invariance \cite{Hosseiny10}
\BEQ \label{2.2}
G = G_0 |t|^{-x}\exp\left[-\frac{{\cal M}}{2} \frac{\vec{r}^2}{t}\right] \;\; ,\;\;  
H = \bigl( H_0 -  G_0 \ln |t|\bigr) \, |t|^{-x} 
\exp\left[-\frac{{\cal M}}{2} \frac{\vec{r}^2}{t}\right] 
\EEQ
subject to the constraint \cite{Bargman54} 
${\cal M}:= {\cal M}_1 = - {\cal M}_2$.\footnote{In order to keep the
physical convention of non-negative masses ${\cal M}\geq 0$, 
one may introduce a `complex conjugate' $\phi^*$ to the
scaling field $\phi$, with ${\cal M}^*=-{\cal M}$. 
In dynamics, co-variant two-point functions are interpreted 
as response functions, written as $R(t,s)=\left\langle \phi(t) \wit{\phi}(s)\right\rangle$ 
in the context of Janssen-de Dominicis theory, where the response field $\wit{\phi}$ 
has a mass $\wit{\cal M}=-{\cal M}$,
see e.g. \cite{Cugliandolo02,Henkel10} for details.\\
Furthermore, the physical relevant equations are {\em stochastic} 
Langevin equations, whose noise terms do break any interesting extended 
dynamical scale-invariance. However, one may identify a `deterministic part' 
which may be Schr\"odinger-invariant, such that the predictions
(\ref{2.2}) remain valid even in the presence of noise \cite{Picone04}. 
This was rediscovered recently under name of
`time-dependent deformation of Schr\"odinger geometry' \cite{Nakayama10}.} 
For the  case of logarithmic conformal Galilean invariance \cite{Henkel10b}
\BEQ \label{2.3}
G = G_0 |t|^{-2x}\exp\left[-2\frac{\vec{\gamma}\cdot\vec{r}}{t}\right] \;\;,\;\;  
H = \bigl( H_0 - 2 G_0 \ln |t|\bigr)\, |t|^{-2x} 
\exp\left[-2\frac{\vec{\gamma}\cdot\vec{r}}{t}\right] 
\EEQ
together with the constraint $\vec{\gamma} :=\vec{\gamma}_1 = \vec{\gamma}_2$. Here, 
$G_0,H_0$ are normalisation constants. 
The presence of the logarithmic terms explain the name of `logarithmic extension'. 


\section{Extension to maximal parabolic sub-algebras} 

Clearly, the results (\ref{2.2},\ref{2.3}) do not contain any information on causality. 
In order to write down the required extension 
of the symmetry algebras, we first 
{\em consider the `mass' parameter $\cal M$ 
as a further variable} (for the moment for the scalar case) 
and write \cite{Giulini96}
\BEQ \label{3.1}
\wht{\phi}(\zeta,t,\vec{r}) := \frac{1}{\sqrt{2\pi\,}} 
\int_{\mathbb{R}} \!\D{\cal M}\: e^{\II{\cal M}\zeta} \phi_{\cal M}(t,\vec{r})
\EEQ
which defines the coordinate $\zeta$ dual to $\cal M$ which we shall consider as a 
`$(-1)^{\rm st}$' coordinate.\footnote{In the
context of string theory and non-relativistic versions of the 
celebrated AdS/CFT correspondence \cite{Maldacena98}, 
an analogous construction is used \cite{Son08,Minic08,Leigh09}, 
with interesting applications to cold atoms \cite{Fuertes09}.}
{}From now on, we concentrate on the case $d=1$ for simplicity. 
The generators of $\mathfrak{sch}(1)$ become 
\BEA
X_n &=& \frac{\II}{2}(n+1)n t^{n-1} {r}^2\partial_{\zeta}
-t^{n+1}\partial_t - \frac{n+1}{2}t^n {r}\partial_r  
- \frac{n+1}{2} x t^n \nonumber \\
Y_m &=& \II \left( m + \demi\right) t^{m-1/2} r\partial_{\zeta} - t^{m+1/2} \partial_j 
\nonumber \\
M_n &=&  \II t^n \partial_{\zeta}  \label{3.2} 
\EEA
The extension to the maximal parabolic sub-algebra 
$\wit{\mathfrak{sch}}(1) = \mathfrak{sch}(1) + \mathbb{C} N$ is achieved by including the
generator
\BEQ \label{3.3}
N := \zeta\partial_{\zeta} - t \partial_t + \xi\,.
\EEQ
 
In order to understand the origin of the constant term $\xi$, 
which in what follows will turn out to play the r\^ole of a second scaling dimension, 
we consider a second representation of the conformal Galilean algebra 
$\mbox{\sc cga}(1) = \left\langle X_{1}, Y_{\pm 1/2}, M_0, V_+, 2X_0-N\right\rangle$, 
see figure~\ref{figab}a. Herein, the generator
$X_1$ takes a slightly generalised form\footnote{The same form of 
$X_1$ also arises in the ageing sub-algebra 
$\mathfrak{age}(1)=\left\langle X_{1,0},Y_{\pm 1/2}, M_0\right\rangle \subset \mathfrak{sch}(1)$. 
Physically, the presence of $\xi$, together with the absence of the time-translations
$X_{-1}=-\partial_t$, leads to distinct exponents $a$ and $a'$ in (\ref{R}).} 
\BEQ
X_1 = {\II} {r}^2\partial_{\zeta}-t^{2}\partial_t - t {r}\partial_r - \left(x+\xi\right) t
\EEQ
along with the new generator
\BEQ
V_+ = -\zeta r \partial_{\zeta} - t r \partial_t 
- \left( \II\zeta t + \frac{r^2}{2} \right)\partial_r - \left( x+\xi\right) r
\EEQ
All other generators are as in (\ref{3.2}). 
One readily verifies that $[V_+, Y_{-1/2}]=2X_0 -N$, 
with the explicitly given forms and this explains the presence of the
constant $\xi$ in (\ref{3.3}). 

The chosen normalisation of the generators is clarified by the commutator 
$[V_+,Y_{1/2}]=X_1$ and the remaining commutators
of $\mbox{\sc cga}(1)$ are promptly verified. 
These generators act as a dynamical symmetry of the Schr\"odinger equation
\BEQ \label{3.6}
{\cal S}\wht{\phi}=0 \;\; , \;\; {\cal S} = 
-2\II\partial_{\zeta}\partial_t - \partial_r^2 
- 2\II \left(x+\xi-\demi\right) t^{-1} \partial_{\zeta}
\EEQ
in the sense that the generators of $\mbox{\sc cga}(1)$ 
map solutions of ${\cal S}\wht{\phi}=0$ onto another solution. 

To check this, it suffices to verify the commutators
\BD
{}[{\cal S}, V_+]=-2r{\cal S} \;\;,\;\;
{}[{\cal S},X_1]=-2t{\cal S} \;\;,\;\; {}[{\cal S},X_0] = -{\cal S} \;\; , \;\; 
{}[{\cal S}, N] = [{\cal S}, Y_{-1/2}] = [{\cal S}, M_0] = 0
\ED
and to recall that $X\wht{\phi}$ with $X\in\mbox{\sc cga}(1)$ 
generates an infinitesimal transformation on the solution $\wht{\phi}$. \hfill q.e.d. 

In general, a {\em standard parabolic sub-algebra} of a simple complex 
Lie algebra is spanned by the Cartan sub-algebra $\mathfrak{h}$ 
and a set of `positive' generators \cite{Knapp86}. 
We illustrate this for the example $B_2$, using figure~\ref{figab}a. The separation between positive
and non-positive generators can be introduced by drawing a straight 
line through the Cartan sub-algebra $\mathfrak{h}$, indicated by the double
point in the centre and then defining all generators who are represented 
by a dot to the right of this line as `positive'. It is well-known
that the Weyl group (which acts on the root diagramme) maps isomorphic sub-algebras onto each other. 
Hence, it is enough to consider the cases when the straight line mentioned
above has a slope between unity and infinity. 
Then one finds the following classification of the non-isomorphic 
maximal standard parabolic sub-algebras of
$B_2$ \cite{Henkel03a}: (i) if the slope is unity, one has $\wit{\mathfrak{sch}}(1)$, 
(ii) for a finite slope larger than unity, one has
$\wit{\mathfrak{age}}(1)=\left\langle X_{0,1},Y_{\pm 1/2}, M_0,N\right\rangle$ and 
(iii) if the slope is infinite, one has $\wit{\mbox{\sc cga}}(1)$. 
 
\section{Dual logarithmic Schr\"odinger-invariance}

We now describe the consequences of logarithmic Schr\"odinger-invariance for the 
`dual' formulation introduced in the previous
section. This representation is constructed from (\ref{3.2}) 
by the formal substitution $x\to \left(\matz{x}{x'}{0}{x}\right)$,
where we explicitly keep the two possibilities $x'=0$ and $x'=1$.  
Only the generators $X_{0,1}$ are modified and now read
\BEA
X_0 &=& -t\partial_t -\demi r \partial_r - \demi \left(\matz{x}{x'}{0}{x}\right) \nonumber  \\
X_1 &=& \frac{\II}{2}r^2\partial_{\zeta} -t^2\partial_t - tr\partial_r  
- t\left(\matz{x}{x'}{0}{x}\right) 
\label{4.1}
\EEA
The co-variant two-point functions, built from quasi-primary scaling operators 
$\left(\vekz{\phi_i}{\psi_i}\right)$ which are characterised
by the values of $x_i$ and $x_i'$, to be studied are
\BEA
\wht{F}(\zeta,t,r) &:=& \left\langle 
\wht{\phi}_1(\zeta_1,t_1,r_1)\wht{\phi}_2(\zeta_2,t_2,r_2)\right\rangle \nonumber \\
\wht{G}_{12}(\zeta,t,r) &:=& \left\langle 
\wht{\phi}_1(\zeta_1,t_1,r_1)\wht{\psi}_2(\zeta_2,t_2,r_2)\right\rangle \nonumber \\
\wht{G}_{21}(\zeta,t,r) &:=& \left\langle 
\wht{\psi}_1(\zeta_1,t_1,r_1)\wht{\phi}_2(\zeta_2,t_2,r_2)\right\rangle \label{4.2} \\
\wht{H}(\zeta,t,r) &:=& \left\langle 
\wht{\psi}_1(\zeta_1,t_1,r_1)\wht{\psi}_2(\zeta_2,t_2,r_2)\right\rangle \nonumber
\EEA
where $\zeta=\zeta_1-\zeta_2$, $t=t_1-t_2$ and $r=r_1-r_2$. 
This form already takes translation-invariance in the three
variables $\zeta,t,r$ into account which in turn follow from the co-variance under 
$M_0,Y_{-1/2},X_{-1}$, respectively.\footnote{Since
the kinetic term of the invariant Schr\"odinger equation (\ref{3.6}) 
reduces to a Laplace operator in a convenient basis, 
the calculations are analogous to those of logarithmic conformal invariance.} Next, we consider
the consequences of co-variance under the Galilei-transformations generated by $Y_{1/2}$. 
For the first of the two-point functions (\ref{4.2}) this implies the differential
equation (called `projective Ward identity' 
in physics\footnote{We prefer to include the terms describing the transformation of the
physical scaling operators right into the generators, 
while many authors only include them into the projective Ward identities. The end
result is the same, the difference corresponds 
to the distinction between active and passive transformations.})
\BEQ \label{4.3}
\left( \II (r_1-r_2)\partial_{\zeta}-(t_1-t_2)\partial_r \right) \wht{F} = 0 
\EEQ
whose general solution (and similarly for the other two-point functions) is
\BEQ \label{4.4}
\wht{F} = \wht{F}(t,u) \;\; , \;\; \wht{G}_{12} = \wht{G}_{12}(t,u) \;\; , \;\; 
\wht{G}_{21} = \wht{G}_{21}(t,u) \;\; , \;\; \wht{H} = \wht{H}(t,u) \;\; ; \;\; 
u := 2\zeta t+\II r^2
\EEQ 
The new specific information of the logarithmic 
representations becomes first evident from dilatation-covariance, generated by $X_0$. 
When taking the previous results (\ref{4.4}) into account, the projective Ward identities become, for the
four distinct functions in (\ref{4.4})
\BEA
\left( -t\partial_t - u\partial_u -\demi (x_1+x_2) \right) 
\wht{F}(t,u) &=& 0 \nonumber \\
\left( -t\partial_t - u\partial_u -\demi (x_1+x_2) \right) 
\wht{G}_{12}(t,u) &=& \frac{x_2'}{2}\wht{F}(t,u)  \nonumber \\
\left( -t\partial_t - u\partial_u -\demi (x_1+x_2) \right) 
\wht{G}_{21}(t,u) &=& \frac{x_1'}{2}\wht{F}(t,u)  \label{4.5} \\
\left( -t\partial_t - u\partial_u -\demi (x_1+x_2) \right) \wht{H}(t,u) &=& 
\frac{x_1'}{2}\wht{G}_{12}(t,u) +  \frac{x_2'}{2}\wht{G}_{21}(t,u) \nonumber 
\EEA
Rather than solving this directly, it is more efficient to use first  
the information coming from the special Schr\"odinger transformations
generated by $X_1$. Applied to the first two-point function $\wht{F}$, 
the use of (\ref{4.3},\ref{4.5}) gives
\BEQ
\left( \frac{\II}{2}r^2\partial_{\zeta}-t^2\partial_t -tr\partial_r -t x_1\right) 
\wht{F}(t,u) = 0
\EEQ
Applying again (\ref{4.5}), we have the system
\BEQ
\left. 
\begin{array}{r} \left( -t\partial_t - u\partial_u -x_1\right) \wht{F} = 0 \\[0.1cm]
\left( -t\partial_t - u\partial_u -(x_1+x_2)/2\right) \wht{F} = 0 
\end{array} \right\}
\Longrightarrow \left( x_1 - x_2\right) \wht{F} = 0 
\EEQ
and we have proven the following\\

\noindent
{\bf Proposition 1:} {\it If $\left(\vekz{\phi}{\psi}\right)$ 
is a quasi-primary scaling operator of logarithmic Schr\"odinger-invariance
with generators (\ref{3.2},\ref{4.1}), 
the two-point function $\wht{F}=\left\langle \wht{\phi}_1\wht{\phi}_2\right\rangle$ 
satisfies one of the following conditions: 
(i) $x_1=x_2$, (ii) $\wht{F}=0$.} 

Now, we consider the mixed two-point functions $\wht{G}_{12}$ and  $\wht{G}_{21}$. 
In complete analogy with the above calculations, we find
\BEQ \label{4.8}
\left. 
\begin{array}{r} \left( -t\partial_t - u\partial_u -x_1\right) \wht{G}_{12} = 0 \\[0.1cm]
\left( -t\partial_t - u\partial_u -(x_1+x_2)/2\right) \wht{G}_{12} - \demi x_2' \wht{F} =0
\end{array} \right\}
\Longrightarrow \left( x_1 - x_2\right) \wht{G}_{12} = x_2' \wht{F} 
\EEQ
and 
\BEQ
\left. 
\begin{array}{r} \left( -t\partial_t - u\partial_u -x_1\right) \wht{G}_{21} = 0 \\[0.1cm]
\left( -t\partial_t - u\partial_u -(x_1+x_2)/2\right) \wht{G}_{21} - \demi x_1' \wht{F} =0
\end{array} \right\}
\Longrightarrow \left( x_1 - x_2\right) \wht{G}_{21} = x_1' \wht{F} 
\EEQ

\noindent
{\bf Proposition 2:} {\it If either $x_2'\ne 0$ and $\wht{G}_{12}\ne 0$ 
or else $x_1'\ne 0$ and $\wht{G}_{21}\ne 0$, then \\
(i) $x := x_1=x_2$ and (ii) $\wht{F}=0$.} 

Obviously, at least one of $\wht{G}_{12}$ or $\wht{G}_{21}$ 
must be non-zero in order to a have non-trivial answer. 
More information is obtained from the last two-point function $\wht{H}$, 
for which covariance under the
generators $X_{0,1}$ implies, using also that $x_1=x_2$
\BEQ \label{4.10}
\left. 
\begin{array}{r} \left( -t\partial_t - u\partial_u -x_1\right) \wht{H} -x_1' \wht{G}_{12}= 0 \\[0.1cm]
\left( -t\partial_t - u\partial_u -(x_1+x_2)/2\right) \wht{H} 
- \demi x_1' \wht{G}_{12}- \demi x_2' \wht{G}_{21} =0
\end{array} \right\}
\Longrightarrow  x_1'\wht{G}_{12} = x_2' \wht{G}_{21} 
\EEQ

Consequently, one must distinguish two essentially distinct cases:
\begin{itemize}
\item[\fbox{$x_1'=x_2'=1$}] We  shall refer to this situation as the {\bf symmetric case}. 
The scaling operators $\left(\vekz{\wht{\phi}_1}{\wht{\psi}_1}\right)$ and 
$\left(\vekz{\wht{\phi}_2}{\wht{\psi}_2}\right)$ are identical. 
Since under the exchange of the two operators, one has
$t\mapsto -t$ and $u\mapsto u$, it follows that $\wht{G}_{12}=\wht{G}(t,u)$ 
and $\wht{G}_{21}=\wht{G}(-t,u)$. 
Because of (\ref{4.10}), the function $\wht{G}(t,u)=\wht{G}(-t,u)$ is symmetric. 
Solving the differential equation (\ref{4.8}), 
we have
\BEQ \label{4.11}
\wht{G}(t,u) = |t|^{-x} \, \wht{g}\left(u |t|^{-1}\right)
\EEQ
where $\wht{g}$ is a differentiable scaling function. Inserting this into 
(\ref{4.10}) and integrating, we find
\BEQ \label{4.12}
\wht{H}(t,u) = |t|^{-x} \left( \wht{h}\left(u |t|^{-1}\right) 
- \ln |t| \;\wht{g}\left( u |t|^{-1}\right) \right) 
\EEQ
Finally, we return to the formulation with fixed masses ${\cal M}_{1,2}$, which gives

\noindent
{\bf Proposition 3:} {\it The co-variant two-point 
functions of the logarithmic representation (\ref{4.1},\ref{3.2})
of $\mathfrak{sch}(1)$ are, with $x := x_1 = x_2$}
\BEA
F(t,r) &=& 0 \nonumber \\
G(t,r) &=& \delta({\cal M}_1+{\cal M}_2)\, |t|^{-x}\: 
\exp\left[-\frac{{\cal M}_1}{2}\frac{r^2}{t}\right] g_0\left( \sign(t), {\cal M}_1\right)
\label{4.13} \\
H(t,r) &=& \delta({\cal M}_1+{\cal M}_2)\, |t|^{-x}\: 
\exp\left[-\frac{{\cal M}_1}{2}\frac{r^2}{t}\right] \left( 
h_0\left( \sign(t), {\cal M}_1\right) - \ln|t|\: g_0\left( \sign(t), {\cal M}_1\right) \right)
\nonumber
\EEA
{\it where $g_0$ and $h_0$ are unspecified functions and $\delta({\cal M})$ is the Dirac distribution.} 

Comparing with the prediction (\ref{2.2}), we can identify $G_0=g_0$ and $H_0=h_0$. 
Notice: logarithmic Schr\"odinger-invariance did {\em not} produce the causality constraint $t>0$~! 

We illustrate the proof of (\ref{4.13}) for $G(t,r)$. 
Using $\zeta=\zeta_1-\zeta_2$, $\eta=\zeta_1+\zeta_2$, we have
\BEA
\lefteqn{G(t,r) = \frac{1}{2\pi} \int_{\mathbb{R}^2} \!\D\zeta_1\D\zeta_2\: 
e^{-\II{\cal M}_1\zeta_1-\II{\cal M}_2\zeta_2}\,
|t|^{-x}\: \wht{g}\left(\frac{2(\zeta_1-\zeta_2) t+\II r^2}{|t|}\right) 
}
\nonumber \\
&=& \frac{1}{4\pi}|t|^{-x}
\int_{\mathbb{R}}\!\D\eta\: e^{-\II({\cal M}_1+{\cal M}_2)\eta/2} \int_{\mathbb{R}}\!\D\zeta\:
e^{-\II({\cal M}_1-{\cal M}_2)\zeta/2}\:\wht{g}\left(2\sign(t) 
\left( \zeta + \frac{\II}{2}\frac{r^2}{\sign(t)\,|t|}\right)\right)
\nonumber \\
&=& \delta({\cal M}_1+{\cal M}_2) |t|^{-x} \int_{\mathbb{R}}\!\D\zeta\: e^{-\II{\cal M}_1 \zeta} \:
\wht{g}\left( 2\, \sign(t) \left( \zeta + \frac{\II}{2}\frac{r^2}{t}\right)\right) \nonumber \\
&=& \delta({\cal M}_1+{\cal M}_2) |t|^{-x} \exp\left[-\frac{{\cal M}_1}{2}\frac{r^2}{t}\right] 
\underbrace{\int_{\mathbb{R}} \!\D\zeta\: 
e^{-\II{\cal M}_1 \zeta}\: \wht{g}\left( 2\, \sign(t) \zeta\right)}_{=\,g_0(\sign(t), {\cal M}_1)}
\nonumber 
\EEA
with a change of variables in the last line and we have also assumed that 
$\wht{g}$ has no singularity `near to' the
real axis which could prevent shifting the contour. $H$ is derived similarly. \hfill q.e.d.

\item[\fbox{$\vekz{x_1'=0}{x_2'=1}$}] This is called the {\bf asymmetric case}. 
The mirror situation $x_1'=1, x_2'=0$ is analogous. 

Now, from (\ref{4.10}) we have $G_{21}=0$. Inserting into and solving 
(\ref{4.8},\ref{4.10}), we have
\BEQ \label{4.14} 
\wht{G}_{12}(t,u) = t^{-x} \,\wht{g}\left(u t^{-1}\right) \;\; , \;\; 
\wht{H}(t,u) = t^{-x} \,\wht{h}\left(u t^{-1}\right)
\EEQ
without any logarithmic term~! Again, no causality condition is produced. 
\end{itemize}

\section{Causality in maximal parabolic sub-algebras} 

In the previous section we had seen that $\mathfrak{sch}(1)$-covariance 
alone is not strong enough to derive the causality condition
$t>0$ for the two-point function. We now show that indeed 
{\em causality is implied if covariance under the maximal parabolic sub-algebra
$\wit{\mathfrak{sch}}(1)$ is required}. In what follows, it will be essential that 
$M_0=\II\partial_{\zeta}$ generates translations in the dual coordinate. In 
consequence, the $M_0$-covariant two-point functions merely depend on $\zeta=\zeta_1-\zeta_2$. 

We begin by extending $N$ to a logarithmic representation by replacing the second scaling dimension 
$\xi$ by a matrix $\Xi=\left( \matz{\xi}{\xi'}{\xi''}{\xi}\right)$ and write
\BEQ \label{5.1}
N = \zeta\partial_{\zeta} - t\partial_t + \left( \matz{\xi}{\xi'}{\xi''}{\xi}\right) \,.
\EEQ

\noindent
{\bf Proposition 4:} {\it One can always arrange in (\ref{5.1}) for $\xi''=0$.} 

Since both $X_0$ and $N$ are in the Cartan sub-algebra of $B_2$, 
see figure~\ref{figab}a, we must have
$[X_0, N] = \demi x'\xi'' \left(\matz{1}{0}{0}{-1}\right) =0$, 
hence $x' \xi'' = 0$. If $x'=0$, one asks whether
$\Xi$ can be diagonalised. If that is so, 
one has the non-interesting case of a pair of non-logarithmic quasi-primary operators. 
If $\Xi$ cannot be diagonalised, it can be brought 
to a Jordan form and one can always arrange for $\xi''=0$. Therefore, we can
set $\xi''=0$ in (\ref{5.1}) without restriction of the generality. One can check the commutators of
$\wit{\mathfrak{sch}}(1)$, notably $[X_1,N]=X_1$. \hfill q.e.d.  

Using the results of section~4, co-variance under $N$ yields
\BEQ \label{5.2}
N \wht{G}_{12}(t,u) = \left( - t\partial_t +\xi_1 + \xi_2 \right) \wht{G}_{12}(t,u) = 0
\EEQ 
Solving this first for $t>0$, this implies $\wht{G}_{12}(t,u)=t^{\xi_1+\xi_2} \wht{\gamma}(u)$. 
Comparison with the scaling form (\ref{4.11}) leads to 
$\wht{G}_{12}=\wht{g}_0 t^{\xi_1+\xi_2} u^{-x-\xi_1-\xi_2}$. 
Together with the results of section~4, and setting $v= u/t$, we have the scaling function
\BEQ \label{5.3}
\wht{g}(v) = \wht{g}_0\, v^{-x-\xi_1-\xi_2} 
\EEQ
where $\wht{g}_0$ is a normalisation constant. 
The last two-point function $\wht{H}$ can be found from
\BEQ \label{5.4}
N \wht{H}(t,u) = \left( -t \partial_t + \xi_1 + \xi_2\right) \wht{H}(t,u) 
+ \xi_1'\wht{G}_{12}(t,u) + \xi_2' \wht{G}_{21}(t,u) = 0 \,.
\EEQ
We now look at the two cases defined in section~4.

\subsection{Symmetric case}
A straightforward calculation gives, using (\ref{4.11},\ref{4.12},\ref{5.3},\ref{5.4})
\BEA
\wht{G}(\zeta,t,r) &=& \wht{g}_0\, |t|^{-x}\: 
\left( \frac{2\zeta t+\II r^2}{|t|}\right)^{-x-\xi_1-\xi_2} \nonumber \\
\wht{H}(\zeta,t,r) &=& |t|^{-x}\: \left( \frac{2\zeta t+\II r^2}{|t|}\right)^{-x-\xi_1-\xi_2} \\
& & \times \left( \wht{h}_0 + \wht{g}_0 (1+\xi_1'+\xi_2') 
\ln\left(\frac{2\zeta t+\II r^2}{|t|}\right) - \wht{g}_0\,\ln|t|\right) \nonumber 
\EEA
where $\wht{g}_0$ and $\wht{h}_0$ are normalisation constants. We can now state the main result.

\noindent
{\bf Theorem.} {\it Quasi-primary scaling operators $\left(\vekz{\phi_i}{\psi_i}\right)$, 
which are scalars under spatial rotations and 
transform co-variantly under a logarithmic representation of the parabolic sub-algebra
$\wit{\mathfrak{sch}}(d)$, are characterised  by the simultaneous Jordan
matrices $\left(\matz{x_i}{x_i'}{0}{x_i}\right)$ and $\left(\matz{\xi_i}{\xi_i'}{0}{\xi_i}\right)$ 
and the masses ${\cal M}_i$. Assume that ${\cal M}_1>0$ and furthermore that
$\demi(x_1+x_2)+\xi_1 +\xi_2>0$. 
If $x_1'=x_2'=1$, the co-variant two-point functions (\ref{2.1}) have the following causal forms}
\BEA
F(t,\vec{r}) &=& 0 \nonumber \\
G(t,\vec{r}) &=& \delta({\cal M}_1+{\cal M}_2)\,\delta_{x_1,x_2}\, \Theta(t)\,t^{-x_1}\: 
\exp\left[ -\frac{{\cal M}_1}{2}\frac{\vec{r}^2}{t}\right] G_0 \label{5.5} \\
H(t,\vec{r}) &=& \delta({\cal M}_1+{\cal M}_2) \,\delta_{x_1,x_2}\,
\Theta(t)\:t^{-x_1} \exp\left[ -\frac{{\cal M}_1}{2}\frac{\vec{r}^2}{t}\right]
\left( H_0 - G_0 \ln t\right) 
\nonumber
\EEA
{\it where $G_0$ and $H_0$ are normalisation constants, $\Theta(t)$ 
is the Heaviside function and $\delta_{a,b}=1$ if $a=b$ 
and zero otherwise.} 

\noindent
The causality statement is contained in the following

\noindent
{\bf Proposition 5:} {\it Let $x>0$, $n$ a non-negative integer 
and consider the integrals, in the limit $\eps\to 0+$}
\BEQ
I_{\pm}^{(n)}(x) := \int_{\mathbb{R}\pm\II \eps} 
\!\D\zeta\: e^{-\II\zeta}\, \zeta^{-x} \ln^n \zeta
\EEQ
{\it Then $I_{-}^{(n)}(x)=0$. There is no simple known expression for $I_{+}^{(n)}(x)$.} 

\begin{figure}[tb]
\centerline{\psfig{figure=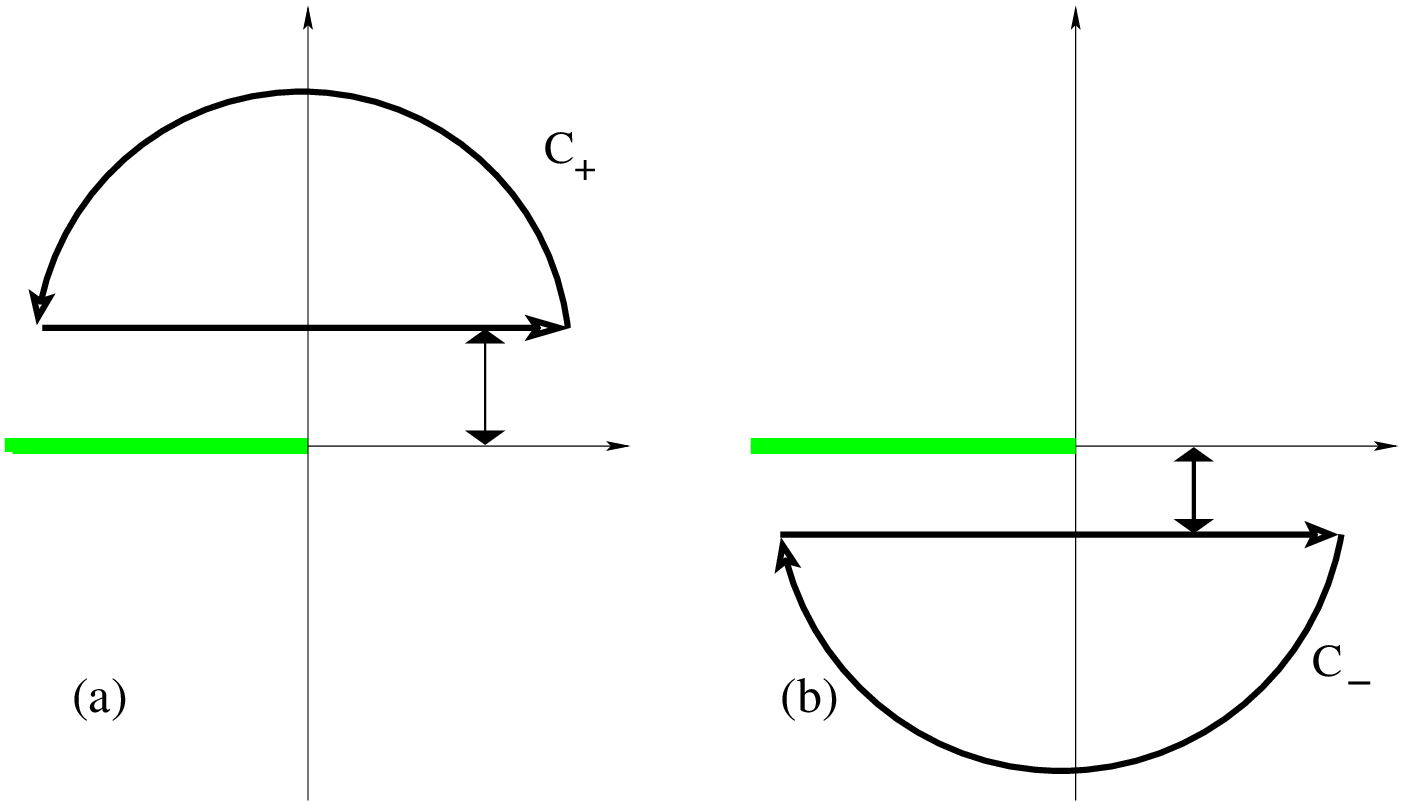,width=4.0in,clip=} }
\caption[figcont]{\label{figcont} Integration contours (a) $C_+$ for $t>0$ and (b) $C_-$ for $t<0$. 
The cut is indicated by the thick line. 
}
\end{figure}

To prove this, consider the contour integrals
\BD
J_{\pm} := \oint_{C_{\pm}} \!\D\zeta\: e^{-\II\zeta}\, \zeta^{-x} \ln^n \zeta
\ED
where the contours $C_{\pm}$ correspond to $t>0$ and $t<0$, respectively, 
as we shall see below and are indicated in figure~\ref{figcont}. 
For $x>0$, the only singularity is the cut along the negative real axis, 
hence $J_{\pm}=0$. We now estimate the contribution of the
lower half-circle, $J_{-,{\rm inf}}$. Setting $\zeta=Re^{\II\theta}$ such that $\ln R>1$, one has
\BD
J_{-,{\rm inf}} = \frac{1}{\II} \int_{0}^{\pi} \!\D\theta\: R^{1-x} 
e^{\II\theta ( x-1) - \II R \cos\theta} e^{-R\sin\theta} 
\left( \ln R e^{-\II \theta} \right)^n 
\ED
Computing the complex logarithm via the binomial theorem, one has the estimate
\BEA
\left| J_{-,{\rm inf}} \right| &\leq & \int_0^{\pi}\!\D \theta\: 
R^{1-x} e^{-R\sin\theta} \sum_{k=0}^n \left(\vekz{n}{k}\right) \ln^{n-k} R 
\: \theta^k \nonumber \\
&\leq & \sum_{k=0}^n \left(\vekz{n}{k}\right) R^{1-x} \left(\ln R\right)^{n-k} \:  \pi^k 
\underbrace{\int_0^{\pi} \!\D\theta\: e^{-R\sin\theta}}_{\leq \pi R^{-1}}
\nonumber \\
&\leq & \pi R^{-x} \left( \pi + \ln R \right)^n \to 0
\nonumber
\EEA
as $R\to\infty$. Hence, since $J_{-} = I_{-}^{(n)}(x) + J_{-,{\rm inf}}=0$, 
the assertion follows. \hfill q.e.d.

In order to prove the theorem, recall first that 
for quasi-primary operators which are scalars under rotations, 
one can always reduce to the case $d=1$. Hence the spatial dependence in (\ref{5.5}) is
a direct consequence of (\ref{4.13}). 
Writing $\xi := \xi_1+\xi_2$, 
we use the physical convention of positive masses ${\cal M}_1>0$ and 
have along the lines of the proof of proposition~3
\BEA
G &=& \delta({\cal M}_1+{\cal M}_2) |t|^{-x} \wht{g}_0 \int_{\mathbb{R}} \!\D\zeta\: 
e^{-\II{\cal M}_1\zeta} \left(2\sign(t)\right)^{-x-\xi} 
\left( \zeta + \frac{\II r^2}{2\sign(t) |t|}\right)^{-x-\xi} \nonumber \\
&=&  \delta({\cal M}_1+{\cal M}_2) \left(2\sign(t)\right)^{-x-\xi} 
{\cal M}_1^{x+\xi-1} |t|^{-x} \wht{g}_0
\underbrace{\int_{\mathbb{R}+\frac{\II {\cal M}_1}{2}\frac{r^2}{t}} 
\!\D\zeta\: e^{-\II\zeta} \zeta^{-x-\xi}}_{I_{\pm}^{(0)}(x+\xi)} 
\; e^{-\frac{{\cal M}_1}{2}\frac{r^2}{t}} 
\nonumber \\
&=& \delta({\cal M}_1+{\cal M}_2) |t|^{-x} 
\underbrace{2^{-x-\xi} {\cal M}_1^{x+\xi-1} \wht{g}_0 I_{+}^{(0)}(x+\xi)}_{=:\, G_0} \:
 e^{-\frac{{\cal M}_1}{2}\frac{r^2}{t}}  \Theta(t)
\nonumber
\EEA
where in the second line we see that for $t>0$ ($t<0$) the contours is slightly above 
(below) the real axis and we need $I_{+}^{(0)}$ 
($I_{-}^{(0)}$). In the last line, 
the statement $I_{-}^{(0)}(x+\xi)=0$ of proposition~5 
was used and expressed by the Heaviside function. 
Similarly, for $H$ we use (\ref{5.5}) and obtain along the same lines
\BEA
\lefteqn{H = \delta({\cal M}_1+{\cal M}_2) |t|^{-x} e^{-\frac{{\cal M}_1}{2}\frac{r^2}{t}}\: 
2^{-x-\xi} {\cal M}_1^{x+\xi-1} \:
\left[ - \wht{g}_0 \ln|t| I_{\pm}^{(0)}(x+\xi) \right. }
\nonumber \\
& &  
\left. + \left(\wht{h}_0 + \wht{g}_0 (1+\xi_1'+\xi_2')\ln(2\sign(t)/{\cal M}_1)\right) 
I_{\pm}^{(0)}(x+\xi) 
 + \wht{g}_0 (1+\xi_1'+\xi_2') I_{\pm}^{(1)}(x+\xi) \right]
\nonumber 
\EEA
and by proposition~5 and defining $H_0$ from the constants in the second line, 
the announced causal form follows. \hfill q.e.d.

\noindent
{\bf Remarks and Generalisations:} (a) Eq. (\ref{5.5}) reproduces the known form (\ref{2.2}) \cite{Hosseiny10}
of logarithmic Schr\"odinger-covariance, but adds
the causality condition $t>0$ described by the extra factor $\Theta(t)$. 
Our derivation generalises earlier causality proofs for 
the non-logarithmic case and under the more strong 
condition $x>0$ \cite{Henkel03a}. \\
(b) For physical applications, recall the form (\ref{R}) of the response function 
$R=\left\langle \phi \wit{\phi}\,\right\rangle$ with
a positive mass ${\cal M}_{\phi}>0$ and a negative mass 
${\cal M}_{\wit{\phi}} = - {\cal M}_{\phi}<0$ such that the `mass conservation' 
following from galilean invariance is accounted for. The response field $\wit{\phi}$ is
associated with the complex conjugate $\phi^*$ in (\ref{3.1}). \\
(c) Since the generator of time-translations $X_{-1}\in\wit{\mathfrak{sch}}(1)$, 
the proven scaling forms (\ref{5.5}) correspond to $a=a'$ in
(\ref{R}). However, the specific form (\ref{3.3},\ref{5.1}) of the generator 
$N$ is already compatible with the more general representations
(or equivalently the Ward identities) required for the 
maximal parabolic extension of the ageing algebra, 
$\wit{\mathfrak{age}}(1)$ \cite{Henkel06a,Henkel10b}. 
Hence the causality arguments presented here explicitly for 
Schr\"odinger-invariance can be directly generalised to ageing-invariance,
including the logarithmic extension. Hence our present results also provide a mathematical 
justification for the sucessful empirical comparison of numerical data of
response functions from critical directed percolation \cite{Henkel10b} 
and the $1D$ KPZ equation \cite{Henkel12} 
with the co-variant two-point function of logarithmic ageing-invariance. \\
(d) Galilei-covariance is an essential assumption. 
While it seems to be well confirmed in many numerical tests of specific models, 
see \cite{Henkel10} and references therein, it is
very difficult to prove formally. Finding such an argument remains an 
important open problem. At present, the nearest one might come to a formal proof is
to consider the models in the dualised form as introduced in section~4. Since therein, 
one trades the phase changes of the usual solution of the 
`Schr\"odinger equation' ${\cal S}\phi=0$ for
a transformation in the dual coordinate $\zeta$, galilean covariance can be checked, 
but of course the procedure modifies seriously the equations under study. 
See \cite{Stoimenov05} for details. \\
(e) The second essential ingredient has been the dualisation with respect to the mass $\cal M$, and
that co-variance under the correspoding generator $M_0 = \II \partial_{\zeta}$ takes the form 
of translation-invariance in the dual coordinate $\zeta$. The importance of this ingredient
can be illustrated by reconsidering briefly the (non-logarithmic) representation 
$\wit{\mbox{\sc cga}}(1)=\left\langle X_1, Y_{\pm 1/2}, D, M_0, V_+, N\right\rangle$ from
section~3, with the dilatation generator 
$D=2X_0-N=-\zeta\partial_{\zeta}-t\partial_t-r\partial_r -\left(x+\xi\right)$. 
In this representation, the only effective scaling dimension apppearing is $x+\xi$, hence the
dual $\mbox{\sc cga}(1)$-covariant two-point function can be read from the 
litt\'erature \cite{Henkel03a,Henkel06b}
\BEA
\left\langle \wht{\phi}_1(\zeta_1,t_1,r_1) \wht{\phi}_2(\zeta_2,t_2,r_2) \right\rangle &=& 
(t_1-t_2)^{-\demi(x_1+\xi_1+x_2+\xi_2)} \left(\frac{t_1}{t_2}\right)^{\demi(x_2+\xi_2-x_1-\xi_1)}
\nonumber \\
& & \times 
f\left( \zeta_1-\zeta_2 + \frac{\II}{2}\frac{(r_1-r_2)^2}{t_1-t_2}\right)
\EEA
Requiring the co-variance $N\left\langle \wht{\phi}_1\wht{\phi}_2\right\rangle = 0$, 
with $N$ given by (\ref{3.3}), leads as before to
$f(u)=\wht{f}_0 u^{-(x_1+3\xi_1+x_2+3\xi_2)/2}$ and transforming back, 
we recover the causality condition $t_1-t_2 >0$, 
provided only that $x_1+3\xi_1+x_2+3\xi_2 >0$. \\
(f) $M_0$ plays the r\^ole of a central extension in the {\em Schr\"odinger} algebra. 
Such a central extension does not exist for $\mbox{\sc cga}(d)$ with $d\ne 2$, 
but we expect that an argument similar to the one used here 
should apply to the exotic central generator $\Theta$ in the {\sc ecga}, after dualisation. 
This should allow, after the identification of the corresponding parabolic sub-algebra, 
to derive causality conditions in this case as well. 
We hope to return to this question in the future. 

\subsection{Asymmetric case} 

Applying the conditions (\ref{5.2},\ref{5.4}) 
to the previously derived scaling forms (\ref{4.14}), we promptly have
\BEQ
\wht{g}(v) = \wht{g}_0 v^{-x-\xi_1-\xi_2} \;\; , \;\;
\wht{h}(v) = v^{-x-\xi_1-\xi_2} \left( \wht{h}_0 - \xi_1' \wht{g}_0 \ln v \right)
\EEQ
Transforming back as before to the situation with fixed masses, 
we obtain under the same conditions as for the main theorem, but now with
$x_1'=0$ and $x_2'=1$, that $F(t,\vec{r})=G_{21}(t,\vec{r})=0$ and the causal, 
but non-logarithmic forms
\BEA
G_{12}(t,\vec{r}) &=& G_0 \delta({\cal M}_1+{\cal M}_2)\,\delta_{x_1,x_2}\, \Theta(t)\,t^{-x_1}\: 
\exp\left[ -\frac{{\cal M}_1}{2}\frac{\vec{r}^2}{t}\right] \nonumber \\
H(t,\vec{r}) &=& H_0 \delta({\cal M}_1+{\cal M}_2) \,\delta_{x_1,x_2}\,
\Theta(t)\:t^{-x_1} \exp\left[ -\frac{{\cal M}_1}{2}\frac{\vec{r}^2}{t}\right] 
\EEA

\noindent
{\small \underline{Note added in proof:} for the representation (\ref{gl:exaltrep}) of the non-exotic {\sc cga}, an 
analogous dualisation and parabolic extension rather shows that 
$\left\langle \phi_1(t)\phi_2(s)\right\rangle = \left\langle \phi_1(s)\phi_2(t)\right\rangle$
is fully symmetric \cite{HenkelStoimenov14}. 
}

\noindent 
{\small 
{\bf Acknowledgement:} It is a pleasure to thank the organisers of the 
7$^{\rm th}$ AGMP conference in Mulhouse 
and especially R. Cherniha for their kind invitation. 
This work was partly supported by the Coll\`ege doctoral franco-allemand Nancy-Leipzig-Coventry 
(Syst\`emes complexes \`a l'\'equilibre et hors \'equilibre) of UFA-DFH.
} 


\end{document}